\documentclass[10pt,letterpaper]{article}
\usepackage{opex3}

\begin{document}

\title{Phase shift spectra of a fiber--microsphere system at the single photon level}

\author{Akira Tanaka$^{\bf 1,2}$, Takeshi Asai$^{\bf 1,2}$, Kiyota Toubaru$^{\bf 1,2}$, Hideaki Takashima$^{\bf 1,2}$, Masazumi Fujiwara$^{\bf 1,2}$, Ryo Okamoto$^{\bf 1,2}$, \\and Shigeki Takeuchi$^{\bf 1,2,*}$}

\address{${}^\textit{1}$Research Institute for Electronic Science, Hokkaido University, Sapporo 001--0020, Japan\\
${}^\textit{2}$The Institute of Scientific and Industrial Research, Osaka University, Mihogaoka 8-1, Ibaraki, Osaka 567-0047, Japan}

\email{${}^\textit{*}$takeuchi@es.hokudai.ac.jp} 


\begin{abstract}
We succeeded in measuring phase shift spectra of a microsphere cavity coupled with a tapered fiber using a weak coherent probe light at the single photon level.
We utilized a tapered fiber with almost no depolarization and constructed a very stable phase shift measurement scheme based on polarization analysis using photon counting.
Using a very weak probe light ($\bar{n}=0.41$), we succeeded in observing the transition in the phase shift spectrum 
between undercoupling and overcoupling (at gap distances of 500 and 100 nm, respectively).
We also used quantum state tomography to obtain a 'purity spectrum'.
Even in the overcoupling regime, the average purity was 0.982$\pm$0.024 (minimum purity: 0.892), 
	suggesting that the coherence of the fiber--microsphere system was well preserved. 
Based on these results, we believe this system is applicable to quantum phase gates using single light  emitters such as diamond nitrogen vacancy centers.
\end{abstract}


\section{Introduction}
\quad Microsphere resonators coupled with tapered optical fibers \cite{MSCTFS1,MSCTFS2} 
	have been attracting interest because of their ultrahigh quality factors \cite{ultrahighq}, 
	small mode volumes \cite{modevolume},	polarization selective coupling \cite{konishi},
	and highly efficient single-spatial-mode input-output \cite{iorelation}. 
Following the pioneering demonstrations of coupling between a microcavity and a tapered fiber \cite{MSCTFS1,MSCTFS2}, 
	applications to lasers \cite{mslaser1,mslaser2,takashima}, 
	biosensors \cite{biosensor}, and nonlinear optics \cite{nonlinear} have been reported. 
When a single light  emitter is deposited on the cavity, 
	the interaction between the light emitter and photons is enhanced due to confinement in  the small mode volume. 
Thus, this system is an ideal testbed for cavity quantum electrodynamics(QED) experiments \cite{takashima2}.
Examples of applications include
	a nonlinear sign shift gate \cite{hofmann,kojima,kimble} for photonic quantum computation \cite{klm} 
	and quantum memory \cite{qrepeater} for long-distance quantum communication. 
We are currently interested in realizing such devices using a fiber--microsphere system 
	with a coupled single light  emitter, e.g. nitrogen-vacancy (NV) centers in diamond \cite{nvcenter}.

To characterize such photonic quantum devices, the probe input light has to be very weak (i.e., single photon level).
It is also essential to analyze the phase change of the probe light to evaluate the coherence properties of quantum devices.
The first step toward performing such evaluations is observing the phase shift of an empty microsphere cavity coupled with a tapered fiber using a probe light at the single photon level \cite{tomita1}. 

\quad A phase shift spectrum was recently obtained
	by interfering the bright coherent signal output from a fiber--microsphere system with a reference light \cite{tomita2}. 
However, it is technically very difficult to stabilize the optical phase 
	between signal and reference lights that have different optical paths,
	especially when using a very faint probe light.

\quad In this letter, we report the measurement of 
	phase shift spectra of a fiber--microsphere system at the single photon level.
To realize stable phase shift measurements of a fiber--microsphere system, 
	we utilized a tapered fiber that has almost no depolarization \cite{konishi, Asai}.
We assume that there are two orthogonal polarization modes, X and Y, in the tapered fiber.
In this case, we can detect the phase shift due to the resonance of the microsphere cavity 
	as a change in the polarization using a probe light that is a superposition of X and Y;
	for example, polarization mode X serves as a reference when only polarization mode Y couples to the cavity mode.
Since neither polarizations exhibit any significant depolarization in the tapered fiber \cite{konishi}, 
	it is not necessary to perform any stabilization, unlike in the previous experiment \cite{tomita2}.
We succeeded in observing a sudden transition in the phase shift spectrum 
	between undercoupling and overcoupling \cite{tomita1} 
	using a very weak probe light with $\bar{n}=0.41$, 
	where $\bar{n}$ is the average number of photons per 10 ns 
	(i.e., typical decay time of a diamond NV center \cite{nvcenter}).

\quad Furthermore, we need to characterize the depolarization of an output photon from a fiber--microsphere system.
For this purpose, we performed frequency-dependent quantum state tomography \cite{james2001} of an output photon
	and derived the purity spectrum, which is important for characterizing the depolarization of the system.
This method involves reconstructing an output state as a density matrix $\hat{\rho} (\omega)$ from the output spectra for three different polarization bases for a polarization input consisting of a combination of two orthogonal polarization modes, X and Y.
Quantum state tomography is able to accurately estimate the coherence even when there are statistical fluctuations, which is significant when a very weak probe light is used. 
From $\hat{\rho}(\omega)$, we obtained purity spectra that indicate the decoherence (i.e. the depolarization) in the fiber--microsphere system. 
The average purity was 0.982$\pm$0.024 (minimum purity: 0.892), indicating that our fiber--microsphere system can maintain coherence and is thus suitable for quantum coherent devices.
Such an observation of the purity spectrum will be essential in future cavity QED experiments.

\section{Experimental Setup}
\begin{figure}[htbp]
\centering\includegraphics[width=12.0cm, height=6.0cm]{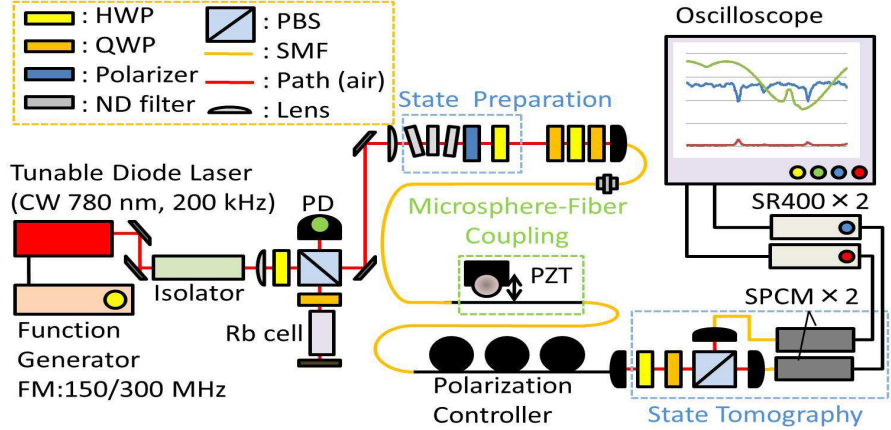}
\caption{Experimental setup. HWP: half wave plate; QWP: quarter wave plate; PBS: polarizing beam splitter; SMF: single mode fiber; PD: photodiode; PZT: piezoelectric transducer; SPCM: single photon counting module.} 
\end{figure}
\quad A tapered fiber was fabricated from a commercial single-mode fiber (Thorlab, 780HP). 
The fiber was heated using a ceramic heater \cite{chiba} and stretched at about $1330{}^{\circ}\mathrm{C}$. 
To generate a large evanescent field strength, 
			a tapered fiber was fabricated with a diameter of 410 nm in the tapered region, 
			as measured by scanning electron microscopy (Keyence, VE9800). 
The total transmittance of the tapered fiber used in the experiment was 30\% (the values of the transmittance given below were compensated by this value). 
A microsphere cavity with a stem \cite{stem} was fabricated by melting the tip of a tapered fiber (fused silica) 
		by focusing a carbon dioxide laser beam (output power: 4--8 W; wavelength: 1.55 $\mu$m) on the tip.
Optical microscopy observations revealed that the microsphere cavity has a diameter of 43.3 $\mu$m.

\quad Fig. 1. shows the measurement setup. 
The probe light is generated by a tunable laser diode (New Focus, Velocity 6312), 
			whose frequency was swept by a function generator about the resonance frequency of the microsphere cavity. 
We set the sweeping range to 150 and 300 MHz; 
			these frequencies correspond to the two coupling conditions.
To calibrate the frequency, 
			we measured the absorption spectrum of a rubidium gas cell.
The laser output was attenuated by three neutral density filters
			to a weak  with $\bar{n}\le1$.
After a polarizing filter, a half-wave plate was used to prepare a combination of X and Y modes at the coupling region  in the tapered fiber.
Two quarter-wave plates and a half-wave plate were used 
			for precompensation of the birefringence in a single-mode fiber.
The probe was then coupled to the single-mode fiber, which was connected to the tapered fiber.
The intensity of the probe light was measured at the connector.
After the tapered fiber, we used an additional polarization controller for postcompensation of the birefringence.
The microsphere was set close to the fiber in the tapered region.
To control the coupling with the cavity, 
			the stem of the microsphere was fixed onto a metal jig of a three-axis piezoelectric transducer (PI, NanoCube)
			to enable the distance between the cavity and the tapered fiber to be varied in 20 nm steps.
The output was sent to the polarization measurement region, which consists of 
			a half-wave plate, a quarter-wave plate, a polarizing beam splitter, 
			two fiber-coupled single-photon counting modules (Parkin Elmer, SPCM-AQR-14) 
			and two photon counters (Stanford Research Systems, SR400). 
The analog outputs of the photon counters (per 1 ms) were sent to an oscilloscope (Tektronix, DPO 4104).
Photon counting spectra, which indicate the coupling of individual polarizations with the microsphere cavity mode, were displayed on the oscilloscope.
The typical photon count was 800counts per 1ms and the dark count of SPCMs were below 0.3counts per 1ms.
The origin of the signal level on the oscilloscope was set when the probe laser was off.

\section{Gap distance dependence of transmittance spectrum}
\quad Fig. 2(a) shows a transmittance spectrum of the tapered fiber obtained when the microsphere cavity was set close to the fiber.
The input light power to the fiber--microsphere system was 10.5 pW, which corresponds to $\bar{n}=0.41$.
The horizontal axis represents the detuning $\Delta f$ of the probe light from the offset frequency $ f_0$ and the vertical axis indicates the transmittance of the input probe light for X polarization.
The offset frequency $f_0$ was set to the frequency at which the minimum was obtained;  the minimum transmittance is approximately 40\% after compensation.
A transmittance of unity was defined as the transmittance obtained when the microsphere was set very far from the tapered fiber.
The dip in the spectrum is due to the coupling \cite{tomita1} between the guided wave mode of the tapered fiber and the whispering gallery mode of the microsphere cavity \cite{phasematched}.
The solid curve in Fig. 2(a) shows a fit of the transmittance spectrum to the transmittance derived from the coupled-mode theory presented in \cite{tomita1}, which is represented by $T(\omega)$ in the following equation:
\begin{equation}
\frac{A_ X(\omega)}{A_{0X}(\omega)}=\sqrt{1-\gamma}\left[\frac{y-x e^{-i\phi}}{1-xye^{-i\phi}}\right]=\sqrt{T(\omega)}e^{-i\theta(\omega)},
\end{equation}
\begin{figure}[htbp]
\centering\includegraphics[width=12.0cm, height=10.0cm]{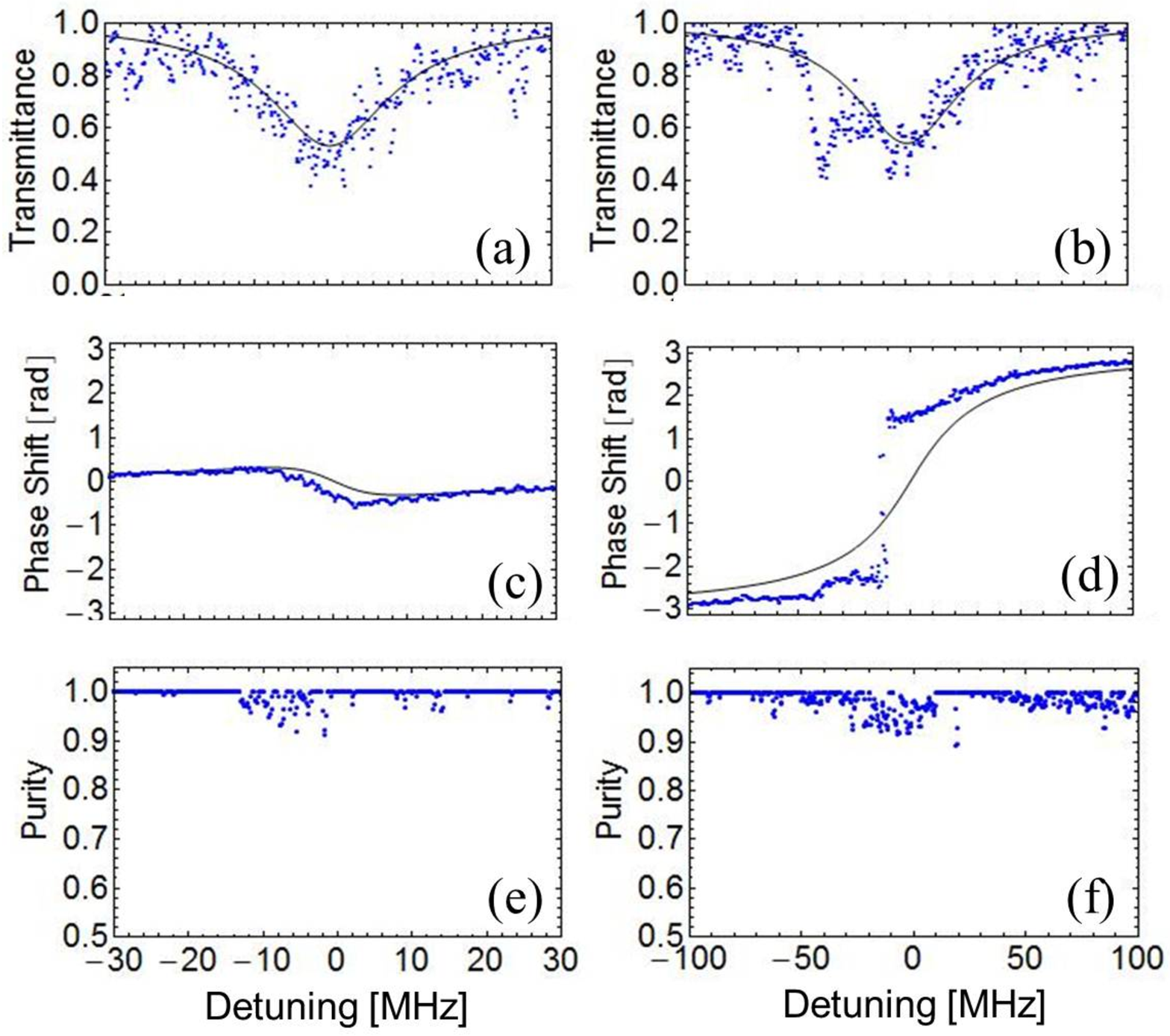}
\caption{(a) Minimum transmittance, (c) phase shift, and (e) purity spectra obtained at a gap distance of 500 nm; (b), (d), and (f) spectra obtained at a gap distance of 100 nm. Solid curves in (a) to (d) are theoretical fits based on coupled-mode theory.} 
\end{figure}
where $A_{0X}(\omega)$ and $A_X(\omega)$ correspond respectively to the complex amplitude of the probe light at an angular frequency $\omega=2\pi (f_0+\Delta f)$ in the tapered fiber in X-polarization before and after fiber--microsphere coupling, $\gamma$ is the total photon loss rate due to coupling, $\phi$ is the round-trip phase in the cavity, and $\theta(\omega)$ is the phase shift of the probe light for X-polarization at $\omega$. 
We defined $x=\sqrt{1-\gamma}e^{-\rho L}$ and $y=\cos{\kappa}$, where $\rho$ is the absorption coefficient of the cavity medium, $L$ is the cavity length, and $\kappa$ is the coupling efficiency from the fiber to the cavity, which is defined as the overlap integral between the two guided wave modes.

\quad Fig. 3 shows the minimum transmittance $T_{min}$ of the probe light 
	as a function of the distance $d$ between the tapered fiber and the microsphere cavity.
	In this experiment, we used a 1 $\mu$W input probe light and a calibrated photodiode, which has placed after the tapered fiber.
The horizontal axis represents the distance $d$
	between the tapered fiber and the surface of the microsphere cavity.
The left vertical axis is the minimum transmittance $T_{min}$ at the resonance frequency 
	(indicated by the red triangles).
Since the spectrum did not change when the microsphere was moved toward the tapered fiber,
	we believe that the microsphere was in contact with the tapered fiber at $d=0$.
As the microsphere  was moved toward the tapered fiber (from approximately 800 nm), 
	$T_{min}$ decreases for $d>300$ nm.
$T_{min}$ has a minimum near $d=d_c\simeq 300$nm.
As $d$ is further reduced, $T_{min}$ tends to increase, as predicted by theory \cite{tomita1}.
The region where $d$ is larger (smaller) than $d_c$ is called the under(over)coupling regime \cite{tomita1,eroded}.
Fig. 2(a) was obtained at $d=500$ nm, where there is undercoupling.
To study the effect of the coupling condition, another transmittance spectrum was obtained at $d=100$ nm (overcoupling) (see Fig. 2(b)).
Note that the scan range of Fig. 2(b) (200MHz) is much larger larger than Fig. 2(a) (60MHz). 
The full width at half maximum (FWHM) of the dip in Fig. 2(b) is three times greater than that in Fig. 2(a), 
	due to the large damping of the cavity mode in the tapered fiber in the overcoupling regime.

\quad For reference, we calculated the quality factor $Q=2\pi f_0/\delta f$ from the FWHM  $\delta f$ of the resonance dip.
$Q$ indicates the degree of confinement of the probe light in the cavity.
The calculated $Q$ for different $d$ is indicated by the green triangles in Fig. 3.
As a result, the quality factor (right vertical axis) decreases continuously from $3.0\times 10^7$ at $d=800$ nm 
	to $1.1\times 10^6$ at $d=0$ nm, as predicted by theory.
The effect of the small parasitic dip at $-50$ MHz in Fig. 2(b) is discussed below.

\begin{figure}[htbp]
\centering\includegraphics[width=10.0cm, height=6.0cm]{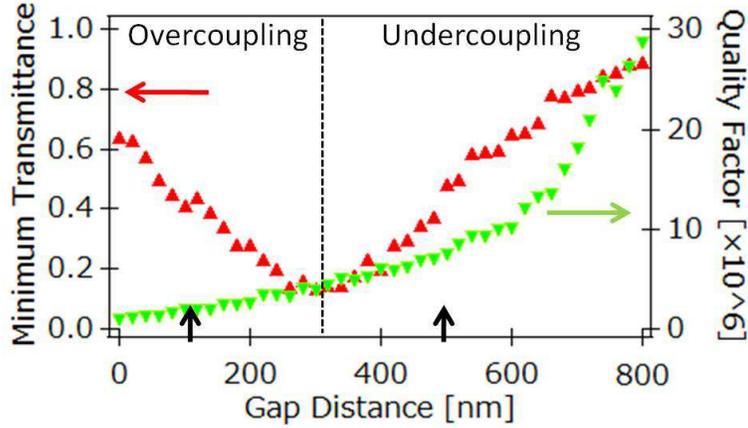}
\caption{Dependences of (red) minimum transmittance and (green) quality factor on gap distance. Black arrows indicate gap distances used to measure the spectra  in Fig. 2.} 
\end{figure}

\section{Phase shift spectrum at the single photon level}
\quad We then measured phase shift spectra for both coupling conditions.
As stated earlier, the phase shift is measured using reference polarization mode Y, 
	which does not couple with the cavity mode; in contrast, X mode couples with the cavity mode, giving a phase shift $\theta_X(\omega)$ described by eq. (1).
We represent the X(Y)-polarized complex amplitude of the input electric field by $A_{0X(0Y)}$.
Here, we consider the case when only $A_{0X}$ couples with the cavity resonance mode
	with a transmittance $T_{all}T_X(\omega)$ and a phase shift $\theta_X(\omega)$.
$T_{all}$ is the total transmittance of both polarization modes.
The output electric field of the fiber--microsphere system then obeys the following transformation:
\begin{eqnarray}
A_{0X}(\omega) \rightarrow A_X(\omega)= \sqrt{T_{all}}\sqrt{T_X(\omega)}e^{i\theta_X(\omega)}A_{0X}(\omega), \nonumber \\
A_{0Y}(\omega) \rightarrow A_Y(\omega)= \sqrt{T_{all}}A_{0Y}(\omega) \ \ \ \ \ \ \ \ \ \ \ \ \ \ \ \ \ \ \ \ \ \ 
\end{eqnarray}

The phase shift $\theta_X(\omega)$ is obtained as follows:
\begin{equation}
\theta_X(\omega)={\rm Tan}^{-1}\left( \frac{S_3(\omega)}{S_2(\omega)}\right)-{\rm Arg} A_{0X} +{\rm Arg} A_{0Y}
\end{equation}
	where $S_2(\omega)=I_P(\omega)-I_M(\omega)$ and $S_3(\omega)=I_R(\omega)-I_L(\omega)$ are Stokes parameters at the frequency $\omega$.
	$I_P(\omega)$ and $I_M(\omega)$ are respectively the intensities of diagonal/antidiagonal polarization modes, and $I_R(\omega)$ and $I_L(\omega)$ are respectively the intensities of right/left circular polarization modes,
	where X and Y are taken to be horizontal and vertical polarization modes.
Note that $I_P(\omega)$ is given by \cite{Yariv}
\begin{eqnarray}
&& I_P(\omega)=\frac{ |A_P(\omega)|^2 }{ 2\eta } \nonumber \\
&& =\frac{1}{2\eta}\left| \frac{A_{X}(\omega)+A_{Y}(\omega)}{\sqrt{2}} \right|^2
\end{eqnarray}
where $\eta=\sqrt{\mu_0/\epsilon_0}$ is vacuum impedance.
Similarly, $I_M(\omega)$, $I_R(\omega)$, and $I_L(\omega)$ are found using $A_M(\omega)=(A_{X}(\omega)-A_{Y}(\omega))/\sqrt{2}$, 
$A_R(\omega)=(A_{X}(\omega)-iA_{Y}(\omega)/\sqrt{2}$, and $A_L(\omega)=(A_{X}(\omega)+iA_{Y}(\omega)/\sqrt{2}$.
\quad Figs. 2(c) and (d) are phase shift spectra obtained for undercoupling (Fig. 2(a)) and overcoupling (Fig. 2(b)), respectively.
The vertical axis indicates the phase shift $\theta_X(\omega)$.
We subtracted a constant phase shift due to the small birefringence of the fiber (0.9 rad in Figs. 2(c) and (d)).
The phase shift asymptotically approaches 0 rad in the highly  detuned region in the undercoupling condition, 
	whereas it approaches $\pm \pi$ in the overcoupling condition ($\pm 2.9 $rad at $\pm 100 $MHz).
Thus, the transition of the phase shift spectrum from undercoupling to overcoupling \cite{tomita2, Asai} is clearly observed using a probe light with $\bar{n}=0.41$.
Note that the parasitic dip at $-50$ MHz in Fig. 2(b) caused a small step-like change at -50MHz in Fig. 2(d).
The change is small (0.3rad) maybe because it is still in the undercoupling condition with small coupling efficiency $\kappa$.

\section{Purity spectrum of the fiber--microsphere system}
\quad We next investigated the depolarization, or dephasing, 
	in this system using quantum state tomography \cite{james2001}.
This method has the advantage that it estimates the most probable state, 
	which is physically meaningful even when there are statistical fluctuations due to finite photon counting, which is critical when $\bar{n}\le1$.  
	This enables the 'purity' to be estimated, which can be used to evaluate the depolarization. 
The density matrix of a single polarization qubit is obtained by measuring 
	Stokes parameters ${\rm S}_0$, ${\rm S}_1$, ${\rm S}_2$, and ${\rm S}_3$ of the output probe light. 
The density matrix $\hat{\rho}$ for the single-photon output state is given by
\begin{equation}
 \hat{\rho}=\frac{1}{2}\left( \hat{I}+\frac{S_1}{S_0}\hat{Z}+\frac{S_2}{S_0}\hat{X}-\frac{S_3}{S_0}\hat{Y} \right).
\end{equation}
Here, $\hat{I}$ is the identity matrix and $\hat{X}$, $\hat{Y}$, and $\hat{Z}$ are the Pauli spin matrices in the basis of
	X- and Y-polarization modes of a single photon.
In quantum state tomography, the density matrix $\hat{\rho}(\omega)$ is estimated using the maximum-likelihood estimation method 
	 with the differential evolution algorithm in Mathematica 7.0 (scaling factor = 1.5)
	 under the constraints (a) ${\rm Tr}\hat{\rho}(\omega)=1$ and (b) $\hat{\rho}(\omega)\ge 0$;
the former constraint states that the total probability is unity
	and the latter constraint states that the probabilities are not negative. 
We performed this tomography for different frequencies $\omega$ and obtained the frequency-dependent $\hat{\rho(\omega)}$.
From $\hat{\rho(\omega)}$, the purity spectrum ${\rm p}(\omega)$ is obtained as follows.
\begin{equation}
{\rm p}(\omega)={\rm Tr}[\hat{\rho}(\omega)^2].
\end{equation}
This purity is  unity when a completely polarized photon experiences no depolarization,
	whereas it is  0.5 when the photon is completely depolarized.
When the absolute detuning from the resonance frequency is relatively large ($|\Delta f| > 20$ MHz in Fig. 2(e)),
	the purity was 0.999 $\pm$ 0.004 (based on an average of 200 points) for undercoupling and 0.998 $\pm$ 0.004 for overcoupling.
This high purity is mainly due to the optimization procedure and the intrinsic nondepolarization of the tapered fiber.
In undercoupling, the purity is 0.992$\pm$0.016 within the bandwidth about the resonance ($|\Delta f|<15$ MHz) (the minimum is 0.913).
In overcoupling, the purity is 0.982$\pm$0.024 within the bandwidth about the resonance ($|\Delta f|<50$ MHz) (the minimum is 0.892.
Note that the purity is high for all detunings.
The possible reasons for the small degradation in the purity spectra about the resonance is considered to be 
	the intrinsic depolarization of the system and spectral jitter.
Thus, using quantum state tomography \cite{james2001}, we have demonstrated 
 	that photons that pass through the fiber--microsphere system experience a tiny depolarization.
Based on this experimental result, which indicates high purity for both coupling conditions, 
	we consider that the fiber--microsphere system is suitable for applications involving coherent quantum devices.
Our recent experiments have demonstrated that  
	it is possible to evaluate photonic quantum devices using, for example, diamond NV centers with a similar setup \cite{takashima2}

\section{Conclusion}
We succeeded in measuring phase shift spectra of a microsphere cavity coupled with a tapered fiber using a weak coherent probe light at the single photon level.
We utilized a tapered fiber with almost no depolarization and constructed a very stable phase shift measurement scheme based on polarization analysis using photon counting.
Using a very weak probe light ($\bar{n}=0.41$), we succeeded in observing the transition in the phase shift spectrum 
between undercoupling and overcoupling (at gap distances of 500 and 100 nm, respectively).
We also used quantum state tomography to obtain a 'purity spectrum'.
Even in the overcoupling regime, the average purity was 0.982$\pm$0.024 (minimum purity: 0.892), 
	suggesting that the coherence of the fiber--microsphere system was well preserved. 
Based on these results, we believe this system is applicable to quantum phase gates using single light  emitters such as diamond nitrogen vacancy centers.

\section*{Acknowledgements}
The current work was supported in part by the program ``R\&D Support Scheme for Funding Selected IT Proposals'' of the Ministry of Public
Management, Home Affairs, Posts and Telecommunications, a Grant-in-Aid from the Japan Society for the Promotion of Science, the 21st
Century COE Program, CREST Project, Japan Science and Technology Agency, JSPS-Grant in Aid for Scientific Research on Innovative areas `Quantum Cyberneticsf,
Funding Program for World-Leading Innovative R\&D on Science and Technology, and Special Coordination Funds for Promoting Science and Technology. 

\begin{thebibliography}{99}
 \bibitem{MSCTFS1}
G. Griffel, S. Arnold, D. Taskent, A. Serpenguzel, J. Connolly, and N. Morris, "Morphology-dependent resonances of a microsphere-optical fiber system," \ol {\bf 21}, 695--697 (1996).
\bibitem{MSCTFS2}
J. C. Knight, G. Cheung, F. Jacques, and T. A. Birks, "Phase-matched excitation of whispering-gallery-mode resonances by a fiber taper," \ol {\bf 22}, 1129--1131 (1997).
\bibitem{ultrahighq} 
M. L. Gorodetsky, A. A. Savchenkov, and V. S. Ilchenko, "Ultimate Q of optical microsphere resonators," \ol {\bf 21}, 453--455 (1996).
\bibitem{modevolume}
J. R. Buck, and H. J. Kimble, "Optimal sizes of dielectric microspheres for cavity QED with strong coupling," \pra {\bf 67}, 033806 (2003).
\bibitem{konishi}
H. Konishi, H. Fujiwara, S. Takeuchi, and K. Sasaki, "Polarization-discriminated spectra of a fiber-microsphere system," \apl {\bf 89}, 121107 (2006).
\bibitem{iorelation}
S. M. Spillane, T. J. Kippenberg, O. J. Painter, and K. J. Vahala, "Ideality in a fiber-taper-coupled microresonator system for application to cavity quantum electrodynamics," \prl {\bf 91}, 043902 (2003).
\bibitem{mslaser1}
M. Cai, O. Painter, K. J. Vahala, and P. C. Sercel, "Fiber-coupled microsphere laser," \ol {\bf 25}, 1430--1432 (2000).
\bibitem{mslaser2}
S. M. Spillane, T. J. Kippenberg, and K. J. Vahala, "Ultralow-threshold Raman laser using a spherical dielectric microcavity," \nat {\bf 415}, 621--623 (2002).
\bibitem{takashima}
H. Takashima, H. Fujiwara, S. Takeuchi, K. Sasaki, and M. Takahashi, "Fiber-microsphere laser with a submicrometer sol-gel silica glass layer codoped with erbium, aluminum, and phosphorus," \apl {\bf 90}, 101103 (2007).
\bibitem{biosensor}
F. Vollmer, S. Arnold, D. Braun, I. Teraoka, and A. Libchaber, "Multiplexed DNA quantification by spectroscopic shift of two microsphere cavities," Biophys. J. {\bf 85}, 1974--1979 (2003).
\bibitem{nonlinear}
I. H. Agha, Y. Okawachi, M. A. Foster, J. E. Sharping, and A. L. Gaeta, "Four-wave-mixing parametric oscillations in dispersion-compensated high-Q silica microspheres," \pra {\bf 76}, 043837 (2007).
\bibitem{takashima2}
H. Takashima, T. Asai, K. Toubaru, M. Fujiwara, K. Sasaki, and S. Takeuchi, "Fiber-microsphere system at cryogenic temperatures toward cavity QED using diamond NV centers," \opex {\bf 18}, 15169--15173 (2010).
\bibitem{hofmann}
H. F. Hofmann, K. Kojima, S. Takeuchi, and K. Sasaki, "Optimized phase switching using a single-atom nonlinearity," J. Opt. B-Quantum S O {\bf 5}, 218--221 (2003).
\bibitem{kojima}
K. Kojima, H. F. Hofmann, S. Takeuchi, and K. Sasaki, "Efficiencies for the single-mode operation of a quantum optical nonlinear shift gate", \pra {\bf 70}, 013810 (2004).
\bibitem{kimble}
Q. A. Turchette, C. J. Hood, W. Lange, H. Mabuchi, and H. J. Kimble, "Measurement of conditional phase-shifts for quantum logic," \prl {\bf 75}, 4710--4713 (1995).
\bibitem{klm}
E. Knill, R. Laflamme, and G. J. Milburn, "A scheme for efficient quantum computation with linear optics," \nat {\bf 409}, 46--52 (2001).
\bibitem{qrepeater}
T. D. Ladd, P. van Loock, K. Nemoto, W. J. Munro, and Y. Yamamoto, "Hybrid quantum repeater based on dispersive CQED interactions between matter qubits and bright coherent light," New J. Phys. {\bf 8}, 184 (2006).
\bibitem{nvcenter}
A. Beveratos, S. Kuhn, R. Brouri, T. Gacoin, J. P. Poizat, and P. Grangier, "Room temperature stable single-photon source," Eur. Phys. J. D {\bf 18}, 191--196 (2002).
\bibitem{tomita1}
K. Totsuka and M. Tomita, "Slow and fast light in a microsphere-optical fiber system," \josab {\bf 23}, 2194--2199 (2006).
\bibitem{tomita2}
M. Tomita, M. Okishio, T. Matsumoto, and K. Totsuka, "Observation of normal and anomalous dispersions in a microsphere taper fiber system," J. Phys. Soc. Jpn. {\bf 78}, 035001 (2009).
\bibitem{Asai}
T. Asai, H. Konishi, H. Takashima, H. Fujiwara, S. Takeuchi, and K. Sasaki, "Optical phase shift observed in a resonance mode of a tapered-fiber coupled with a microsphere resonator", in \textit{Meeting Abstracts of the Phys. Soc. of Japan}, (Academic, Osaka, Japan, 2008) {\bf 63}, 23pQD-13, pp. 152.
\bibitem{james2001}
D. F. V. James, P. G. Kwiat, W. J. Munro, and A. G. White, "Measurement of qubits," \pra {\bf 64}, 052312 (2001).
\bibitem{chiba}
A. Chiba, H. Fujiwara, J. Hotta, S. Takeuchi, and K. Sasaki, "Fano resonance in a multimode tapered fiber coupled with a microspherical cavity," \apl 86, 261106 (2005).
\bibitem{stem}
L. Collot, V. Lefevreseguin, M. Brune, J. M. Raimond, and S. Haroche, "Very high-Q whispering-gallery mode resonances observed on fused-silica microspheres," Europhys. Lett. {\bf 23}, 327--334 (1993).
\bibitem{phasematched}
J. C. Knight, G. Cheung, F. Jacques, and T. A. Birks, "Phase-matched excitation of whispering-gallery-mode resonances by a fiber taper," \ol {\bf 22}, 1129--1131 (1997).
\bibitem{eroded}
N. Dubreuil, J. C. Knight, D. K. Leventhal, V. Sandoghdar, J. Hare, and V. Lefevre, "Eroded monomode optical-fiber for whispering-gallery mode excitation in fused-silica microspheres," \ol {\bf 20}, 813--815 (1995).
\bibitem{Yariv}
A. Yariv, "Optical Electronics in Modern Communications", pp. 12 (Oxford, New York, 1997).
\end{thebibliography}
\end{document}